# Entropic Measure of Time, and Gas Expansion in Vacuum

**Leonid M. Martyushev [1,2,\*], Evgenii V. Shaiapin [1]**

[1] Ural Federal University, 19 Mira St., Ekaterinburg, 620002 Russia; shayapin@mail.ru
[2] Institute of Industrial Ecology, Russian Academy of Sciences, 20 S. Kovalevskaya St., Ekaterinburg, 620219 Russia
[\*] Correspondence: LeonidMartyushev@gmail.com; Tel.: +7-922-22-77425

**Abstract:** The study considers advantages of the introduced measure of time based on the entropy change under irreversible processes (entropy production). Using the example of non-equilibrium expansion of an ideal gas in vacuum, such a measure is introduced with the help of Boltzmann's classic entropy. It is shown that, in the general case, this measure of time proves to be nonlinearly related to the reference measure assumed uniform by convention. The connection between this result and the results of other authors investigating measure of time in some biological and cosmological problems is noted.

**Keywords:** entropy production; intrinsic and chronological times; ideal gas expansion.

## 1. Introduction

The second law of thermodynamics together with the following concept of entropy had a tremendous effect on physics and natural science in general. All these – introduction of an absolute temperature scale, restriction of the conversion efficiency of heat engines, discovery of various regularities under nonequilibrium processes – and many other things originated as crucial corollaries of the second law [1]. The increase of entropy in an isolated system enabled to introduce a physical quantity characterizing irreversibility, an essential property of the world around us that was not previously covered by the laws of mechanics and electromagnetics.

The irreversibility and directionality from the past to the future is a basic property of another physical quantity: time. The problem of time and the properties thereof are addressed by an enormous number of studies, especially philosophic ones. Physicists investigate this quantity much more rarely; however, this area has a number of good reviews (see, for instance, [2-6]). Based on these investigations, we can identify a number of properties that are common for time and entropy. Thus, in addition to directionality, they both are related to variability and depend on an observer. The latter property is more often attributed to Boltzmann's and the subsequent information and statistical generalization of entropy as well as to the most wide-spread approach in the present post-Einstein age: relational approach to time. Furthermore, both quantities are considered among fundamental and most difficult-to-study concepts characterizing the surrounding world.

As regards the noted affinity, there is a number of papers (see, for instance, [7]) where the measurement of time flow is directly associated with the measurement of entropy change (more specifically, with irreversible change, i.e. entropy production). This use of entropy production as a metric of time has a considerable advantage as compared with others. Namely, the traditional methods of introducing a metric of time are based on astronomic, mechanic, electromagnetic, and quantum optical regularities. However, according to the contemporary understanding, they are seen as essentially reversible phenomena. Consequently, these regularities are unreasonable from the fundamental perspective, as a *basic measure* for the essentially irreversible quantity (the so-called arrow of time). Universality is another advantage of the entropy-based time. Indeed, the statistical introduction of entropy through a number of microstates enabling a system's state allows finding entropy and, therefore, time for any systems (including imaginary and modelled ones). As opposed to the entropic measure of time, the traditional measures of time are based on some specific phenomenon that a system may lack in the general case (particularly, it is easy to assume a hypothetic world of electrically neutral classic particles where electromagnetic and quantum optical measures of time cannot be used). Thus, the entropy method of introducing a metric of time flow appears to be as universal as the traditional method of introducing a space metric. Moreover, what is very important, this method is essentially different from the methods associated only with spatial measurements (as distinct, for example, from the employed astronomic and mechanical methods for introducing measure of time). In addition, the fact that the time considered herein is directly related to a system's dissipative processes and is, therefore, strictly individual for every system involved represents a major feature of this time and its fundamental difference from an absolute Newtonian time.

In spite of the reasons above, the question about relations between entropy and measure of time is still poorly developed in the literature. Considerable discussions of this topic from the qualitative perspective can be more often found in philosophic and popular-science editions (see, for instance, [2-5,8]), while quantitative, relatively rigorous investigations of such relations and their possible corollaries are obviously insufficient so far (see, for instance, [9-12]).

The effort in this area was made in [13]. The metric of time for a developing system $\tau$ was introduced as directly proportional to the specific production of thermodynamic entropy. The found logarithmic relation of $\tau$ to the reference (uniform) time $t$ used by the observer external to the system is an interesting result of such a consideration. A drawback of this research is that the class of systems for which the provided approach is directly suitable is restricted by the postulates of local nonequilibrium thermodynamics. As is known, these restrictions depend on the possibility to consider a system in a local equilibrium state. The class of nonequilibrium systems which provide for such an opportunity is very wide (it includes the majority of practically important systems that we encounter). However, there still are systems for which such a consideration is inapplicable [14]. As a result, for these systems it is necessary to find a method of introducing the entropy-based time and analyze the possible corollaries thereof. This subject is the purpose of the present study.

**2. Model**

In order to achieve the stated purpose, we deliberately chose to study a very complicated case in *terms of thermodynamic description*: expansion of an ideal gas in vacuum, i.e. motion of identical particles that can interact rather rarely only by absolutely elastic collisions. Initially, the gas occupies a certain volume where molecules' locations and velocities are unknown and reasonably arbitrary. The direction of motion is strictly radial. Then the gas starts expanding; during this process, the volume occupied by its molecules increases while the initial distribution of particle velocities, according to the model involved, remains the same. Obviously, any stage of expansion in such a system is not in a thermodynamic equilibrium (including the local one), and thermodynamic equilibrium characteristics such as temperature and the like cannot be introduced. For illustrative purposes, we will consider a one-dimensional case (Fig.1). We note that a three-dimensional case is absolutely similar and will be studied below. Let us assume a number of particles $N$. We have an observer inside the expanding gas having the above properties. The observer would naturally choose a size of one of the system's particles as a length scale. Then other spatial sizes of the problem are normalized to this length scale, i.e. considered dimensionless. These include initial and current sizes of the regions occupied by the gas: $r_0$ and $r$, respectively (see Fig.1). The observer has no clock (timepiece). However, the purpose of the observer is to introduce measure of time based on the properties of the system of particles around him. Using the above mentioned properties of measure of time and following this paper's purpose, the observer wants to use the entropy method. The statistical entropy that can be introduced for arbitrary, including essentially nonequilibrium, systems [15] would be naturally chosen. For the problem under consideration, the entropy change equals the entropy production (no entropy flows through the system's boundaries). Since only spatial changes occur in the system, the observer divides the system's current size $r$ into $G$ similar cells having the size $\Delta$ (Fig. 1). This size remains unchanged and, additionally, can contain up to $N$ particles. It is obvious that, at different system-observation instants, $G$ varies and equals $2r/\Delta$. The size of the cell selected by the observer depends on the degree of detail in which the observer prefers to describe the system. Based on the information which is known by the observer (or, as we should rather say, a total lack of information about the system, including details of its initial state), all possible distributions of particles by cells should be considered equally probable. This probability is inversely proportional to the number of such distributions $\Omega$. This number by which $N$ identical particles may be arranged by $G$ cells with an arbitrary number of particles per cell is well-known [15]:

$$\Omega = \frac{(N+G-1)!}{N!(G-1)!}. \tag{1}$$

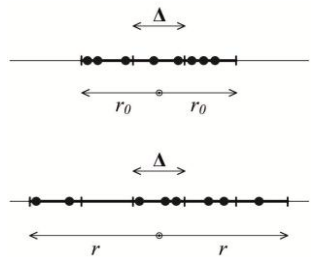

**Figure 1.** The simplest model of identical particles spreading out in vacuum considered herein. The initial state and one of the following states are shown. All symbols in the figure are explained in the text hereof.

A similar formula is traditionally used in the literature for describing boson distributions. The difference between the two formulae lies only in their interpretations. Thus, in the case of bosons, cells are primarily associated with energetic rather than spatial states and the number of cells is usually considered constant as opposed to variable in our case.

## 3. Entropic measure of time

As noted above, the observer uses such a measure of time that the time change $d\tau$ is equal (accurate up to a constant multiplier) to the irreversible change of entropy per particle $dS_{ir}$ (entropy production density):

$$d\tau = dS_{ir}/N. \tag{2}$$

In Eq.(2), the Boltzmann entropy (which in some known cases is reduced to thermodynamic entropy) is used and the entropy changes due to irreversible processes inside the system rather than a flow across the system's boundaries. It should be additionally emphasized that the entropic measure (2) has meaning only if irreversible processes take place in the system. In the case of equilibrium, and for hypothetic reversible processes, it loses its meaning.

The Boltzmann entropy $S$ for the system under consideration is defined (accurate up to a constant dimensional multiplier) by [15]:

$$S = \ln \Omega. \tag{3}$$

Assuming that $N \gg 1$ and $G \gg 1$, based on Eqs. (1) and (3), using Stirling's approximation, we obtain:

$$S = (N+G)\ln(N+G) - G\ln G - N\ln N. \tag{4}$$

As a result, for entropy per particle we have:

$$S/N = \left(1+\frac{G}{N}\right)\ln\left(1+\frac{G}{N}\right) - \frac{G}{N}\ln\left(\frac{G}{N}\right). \tag{5}$$

For the gas expansion problem at hand, obviously $dS_{ir} = dS$. Therefore, using Eq. (2), the following can be written, accurately up to an additive constant:

$$\tau \equiv S/N = \left(1+\frac{G}{N}\right)\ln\left(1+\frac{G}{N}\right) - \frac{G}{N}\ln\left(\frac{G}{N}\right). \tag{6}$$

The measure of time thus introduced has all the necessary properties. So, for the preset $N$, it increases monotonically under the gas's irreversible expansion which is always accompanied by the increase of $G$. Since $G = 2r/\Delta$, the introduced measure is immediately related to the changes occurring in the system and depends on the observer dividing the space available to the system into similar cells. Furthermore, the introduced measure is rather universal: it requires only an observer (selecting Δ), a developing system, and the possibility of introducing measure of distance.

It is important to establish relations between the introduced measure of time (6) and the time $t$ assumed as reference uniform time. For this purpose, we shall consider an observer positioned outside the system at hand and having such a reference clock. This observer has all the information about the system like the first observer, i.e. knowledge of the quantities $r_0$, $r$, $N$, $G$, and $Δ$. The outside observer will calculate the system's specific entropy in exactly the same manner as in Eq. (5). However, originally having a clock, the observer can introduce the velocity of motion $υ=dr/dt$. Therefore, for the considered one-dimensional motion, $G = 2(r_0 + υ\, t)/Δ$. As a result, according to Eq.(6),

$$\tau = \left(1 + \frac{2(r_o + υ \cdot t)}{NΔ}\right)\ln\left(1 + \frac{2(r_o + υ \cdot t)}{NΔ}\right) - \frac{2(r_0 + υ \cdot t)}{NΔ}\ln\left(\frac{2(x_o + υ \cdot t)}{NΔ}\right). \tag{7}$$

Let us introduce the following notation: $\alpha_1 = \dfrac{2r_0}{NΔ}$, $t_1 = \dfrac{NΔ}{2υ}$. Then we obtain

$$\tau = (1 + \alpha_1 + t/t_1)\ln(1 + \alpha_1 + t/t_1) - (\alpha_1 + t/t_1)\ln(\alpha_1 + t/t_1). \tag{8}$$

Using Eq. (8), we will consider two limits:
Let $t \to 0$, then

$$\tau \propto \tau_1 + \xi_1 t/t_1, \tag{9}$$

where $\tau_1 = (1 + \alpha_1)\ln(1 + \alpha_1) - \alpha_1 \ln \alpha_1$, $\xi_1 = \ln(1 + 1/\alpha_1)$.

Let $t \to \infty$, then

$$\tau \propto \ln(t/t_1). \tag{10}$$

The obtained formulae (8)–(10) indicate that, for the general case, the time $\tau$ introduced by the inside observer based on the calculations of entropy is nonuniform (nonlinear) with respect to $t$ (Fig.2). The relation of the two scales appears to be linear only in the very beginning of expansion (see Eq.(9)). It may seem that such a property of the entropic measure of time is its drawback. Indeed, it is traditionally assumed in physics that time flows uniformly for velocities much smaller than the velocity of light and for relatively small masses (i.e. outside the scope of the special and general theory of relativity). However, such an assumption is not based on any law; moreover, it cannot be proved neither logically nor empirically. In particular, this issue was addressed by H. Poincaré [16,17]. Let us here quote his paper [16]: *"We have not a direct intuition of the equality of two intervals of time. The persons who believe they possess this intuition are dupes of an illusion. When I say, from noon to one the same time passes as from two to three, what meaning has this affirmation? The least reflection shows that by itself it has none at all. It will only have that which I choose to give it, by a definition which will certainly possess a certain degree of arbitrariness."* And another quotation of his [16]: *"…there is not one way of measuring time more true than another; that which is generally adopted is only more convenient. Of two watches, we have no right to say that the one goes true, the other wrong; we can only say that it is advantageous to conform to the indications of the first."* One has to remember that we have assumed the existence of the uniform scale of $t$ that can be used as reference only by convention. In reality (for instance, from the standpoint of some third observer), the clock of $t$ of the outside observer considered herein may be even more "nonuniform" than that of the inside observer.

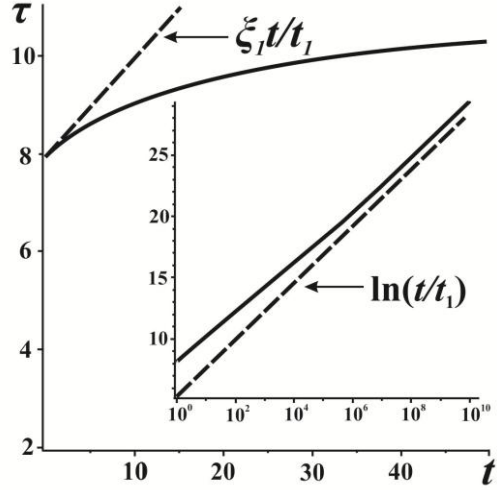

**Figure 2.** Dependence of the entropic time $\tau$ introduced by the observer inside the system on the time chosen by the outside observer as the uniform time $t$. The dashed lines show limit approximations Eq.(11) at small and large times. $t_1 = 0.005$, $\alpha_1 = 1000$.

Such nonlinearity leads to interesting kinematical corollaries. Indeed, according to the outside observer, the system's boundary expands at a constant velocity $\upsilon$, the velocity of motion of the originally fastest particles. However, from the perspective of the inside observer, such constancy of velocity will be observed only in the beginning, and at considerably long times the velocity of motion will grow exponentially. Indeed, according to the inside observer, the velocity $dr/d\tau$ can be written, using Eqs. (9,10), as:

$$\frac{dr}{d\tau} = \frac{dr}{dt}\frac{dt}{d\tau} = \upsilon \frac{dt}{d\tau} = \begin{cases} \dfrac{N\Delta}{2\xi_1} & t,\tau \to 0 \\[2mm] \dfrac{N\Delta}{2} e^\tau & t,\tau \to \infty \end{cases}. \tag{11}$$

As is seen, at some moment of observation, according to the measurements of the inside observer using Newton's dynamics to describe the motion, the particles will start spreading out with an acceleration as if under the influence of a "mysterious" (or "dark") force which arises and becomes stronger with time.

Let us make two important notes.

1. Previously, the one-dimensional case was considered. It can be easily extended to a three-dimensional one. The gas expands radially with the spherical symmetry. Initially, it is contained in a sphere having the radius $r_0$ and then occupies ever larger spheres of the radius $r$. As before, let us write the number of cells and the number of particles as $G$ and $N$, respectively. As in the case above, we will assume the volume of cells constant and designate it as $\Delta^3$. Given the symmetry of the problem, the shape of the cells can be selected as spherical layers of some thickness centered on the system's point of symmetry. Obviously, the thickness of these layers is to decrease in inverse proportion to $r^2$. For such a formulation, $G = 4\pi(r/\Delta)^3/3$. It is evident that the expression for the time measured by the inside observer using the variables $G$ and $N$ will remain the same as Eq. (6). However, the relation of the inside observer's time and the reference time has the form:

$$\tau = \left(1 + (\alpha_3 + t/t_3)^3\right)\ln\left(1 + (\alpha_3 + t/t_3)^3\right) - 3(\alpha_3 + t/t_3)^3 \ln(\alpha_3 + t/t_3). \tag{12}$$

where $\alpha_3 = (4\pi/3N)^{1/3} r_0/\Delta$, $t_3 = (3N/4\pi)^{1/3}\Delta/\upsilon$. Two limit cases have the form:

$$\tau \propto \tau_3 + \xi_3 t / t_3, \text{ with } t \to 0, \quad (13)$$

where $\tau_3 = (1+\alpha_3^3)\ln(1+\alpha_3^3) - \alpha_3^3 \ln(\alpha_3^3)$, $\xi_3 = 3\alpha_3^2 \ln(1+1/\alpha_3^3)$.

$$\tau \propto 3\ln(t/t_3), \text{ with } t \to \infty. \quad (14)$$

Thus, as is seen from the given formulae, the three-dimensional case has no fundamental differences from the linear one.

2. Previously, we have considered a strongly nonequilibrium case of gas expansion in vacuum. This case cannot be investigated using the methods of classical thermodynamics. It can be shown, nevertheless, that, for a number of adjacent problems, results similar to the ones above can be obtained thermodynamically using a number of restrictions. We will consider some initial equilibrium state of an ideal gas. The gas adiabatically expands from it in vacuum and reaches another equilibrium state while changing the volume $V$. Obviously, this process is irreversible and the gas does no work. According to the adiabatic nature of the process and the first law of thermodynamics, the temperature of an ideal gas is to remain unchanged during expansion. Let us replace a real irreversible process with a hypothetic isothermal equilibrium process having identical initial and final states. The changes of thermodynamic entropy $S_t$ for the two processes are the same and it is easy to show [1,15] that the change of entropy in the case at hand is equal to

$$dS_t = \nu R \, dV / V, \quad (15)$$

where $R$ is the universal gas constant and $\nu$ is the number of gas moles.

As before, by introducing the inside observer's time as directly proportional to entropy per particle, we obtain

$$d\tau \propto dS_t / \nu, \quad (16)$$

or, with the accuracy up to an additive constant and multipliers,

$$\tau = \ln V. \quad (17)$$

For the outside observer (having the reference uniform clock of $t$), the gas volume during expansion is described by a power-law dependence on $t$. By inserting this law into the last formula, a logarithmic relation between the two times, similar to Eq.(14), is obtained.

**4. Conclusion**

The present study develops a formerly proposed relation between measure of time and entropy. The thermodynamic entropy previously used for this purpose in the case of locally-equilibrium processes is replaced with the Boltzmann entropy suitable for describing arbitrary nonequilibrium systems. For the simplest model of ideal-gas expansion, we obtain the entropic measure of time related to the spatial disorder and directed towards its increase.

It was not an intention hereof to answer metaphysical questions about the cause of variability in the world, the nature of time, its reality, and the like. We considered a purely physical problem of the most consistent and universal, from the theoretical point of view, introduction of measure of time. Such a rigorous operational approach to the introduction of measures (particularly, measure of time) of the basic physical quantities is extremely important for the foundation of physics and its further development. It was mentioned many times by P. W. Bridgman [18], E.A. Milne [19], et al.

Importantly, it is the existence of a special function of state, entropy, (i.e. one of the formulations of the second law of thermodynamics) that allowed introducing an absolute temperature scale for a thermodynamic system in a thermal equilibrium. As a result, an objective measure of thermal-motion intensity appeared in science [1]. The present study connects the change of entropy under irreversible processes with the possibility to consistently introduce measure of time. The noted relations between entropy, on the one side, and temperature/time, on the other side, may prove to be very meaningful and require an additional research.

The crucial result hereof is that a nonlinear relation between the time scale used by an observer and the one chosen as reference is established for the general case. We believe that a logarithmic relation observed at relatively long times are especially important here (see Eq.(14)). To be specific, such a relation between intrinsic (developmental) and chronological (astronomical) times during growth and development of biological systems was independently proposed before [13,20,21]. In particular, the papers [13, 21] provide a theoretical grounding for the universal power-law relationship between a developing system's mass and astronomical time. This relationship is confirmed by the available empirical data of the growth in biological and crystallization systems. Then the change of mass in the system is connected with its entropy production which, similarly to the approach hereof, is directly related to the internal (biological) time in the system. Additionally, such a logarithmic relation between the two scales of time and its potential significance for cosmology was also mentioned in 1937-1950s by E.A. Milne (see, for instance, [19]). It is interesting that E.A. Milne obtained this relation only kinematically while bringing his laws of motion obtained for the so-called fundamental particles (galactic nuclei) into consistency with the traditional Newton laws. The fundamental particles formed a basis of his cosmological model of the world. These particles spread out uniformly in different directions at different velocities (largely like the expansion of an ideal gas in vacuum considered herein). Milne shows that the uniform motion of the fundamental particles occurs for the time $t$ (a universal time related to the atomic clock), whereas $\tau$ (which the logarithmically associated with $t$) is a time introduced by the researcher on the basis of astronomic observations. Such use of the two times allowed E.A. Milne to build an original theory that had a great influence on the development of modern cosmology.

The parallels mentioned herein obviously demonstrate a common nature and a close connection between the measure of time considered in the present paper and the measures of time used both in the problems of cosmology to describe the origin and development of the Universe and in the problems of biology to describe the birth and growth of living beings.

Despite the result obtained herein with respect to the logarithmic relation between the measures of time and its mentioned connection with the results of other studies, the question of universality of such a relation requires further serious analysis and, presently, generalizations (in particular, regarding cosmological problems) must be avoided.

**Acknowledgments:** The authors express their special thanks to Prof. Vladimir D. Seleznev for discussion and unfailing interest to the topic considered herein. His death on October 12, 2015, became a tremendous personal tragedy for us. His memory will always be with us. We dedicate this research on entropy and time to him. Vladimir D. Seleznev contributed the most of his Time on this earth to the study of entropy and his ideas will dwell with us for a very long time.

The reported study was partially funded by RFBR according to the research project No.16-31-00255 мол_а and one of the authors (E.V.S) expresses his gratitude for it.

**Author Contributions:** Leonid M. Martyushev proposed all the basic ideas of the research including the problem of calculation of the entropic measure of time for an ideal gas expanding in vacuum. He also wrote the basic text of the paper. Evgenii V. Shaiapin carried out all the necessary calculations and prepared the manuscript for publication. Both authors analyzed and discussed the results, read and approved the final version of the paper.

**Conflicts of Interest:** The authors declare no conflict of interest.